\documentclass[nofootinbib]{revtex4}
\usepackage{graphicx}
\usepackage{latexsym}
\def\be{\begin{equation}}
\def\ee{\end{equation}}
\def\bea{\begin{eqnarray}}
\def\eea{\end{eqnarray}}

\begin{document}
\title{Curvaton Reheating in Non-oscillatory Inflationary Models}
\author{Bo Feng}
\email{fengbo@mail.ihep.ac.cn}
\author{Mingzhe Li}
\email{limz@mail.ihep.ac.cn}
\affiliation{Institute of High Energy Physics, Chinese
Academy of Sciences, P.O. Box 918-4, Beijing 100039, People's Republic of China}

\begin{abstract}
In non-oscillatory (NO) inflationary models, the reheating
mechanism was usually based on gravitational particle production or the mechanism of instant preheating.
In this paper we introduce the curvaton mechanism
into NO models to reheat the universe and generate the curvature perturbations.
Specifically we consider the Peebles-Vilenkin quintessential
inflation model, where the reheating temperature can be extended
from $1$MeV to $10^{13}$GeV.
\end{abstract}

\maketitle

\hskip 1.6cm
PACS number(s): 98.80.Cq

Inflarionary universe \cite{guth} has solved many problems of the Standard Hot Big-Bang
theory, e.g., flatness problem, horizon problem, etc.
In addition, it has provided a causal interpretation of the origin of
the observed anisotropy of the cosmic microwave background (CMB) and the distribution of
large scale structure (LSS).
In the usual inflation models, the acceleration is driven by the
potential of a scalar field $\phi$ (inflaton) and its quantum fluctuations in this epoch generate the
density perturbations seeding the structure formations at late time.
To date, the accumulating observational data, especially those
from the CMB observation of WMAP satellite \cite{wmap1,wmap2,wmap3} indicate the power
spectrum of the primordial density perturbations is nearly
scale-invariant, adiabatic, and Gaussian--just as predicted by the
single-field inflation in the context of ``slow-roll''.
Generally, after inflation the inflaton field will oscillate about the minimum
of its potential and eventually decay to produce almost all elementary particles populating the
universe. This process is called reheating.

However, there exist some inflationary models \cite{no,ford,vilenkin,joyce}, named NO models by
the authors of Ref. \cite{kofman}, in which
the minimum of the inflaton potential exist in the infinity.
In these models the inflaton field would not oscillate after inflation and
the standard reheating mechanism could not work.
Conventionally one turns to gravitational particle
production \cite{ford}. However, the authors of Ref. \cite{kofman}
pointed out that this mechanism is very inefficient. For the resolution to this problem, they
proposed to use ``instant preheating'' \cite{felder} by
introducing an interaction $g^2\phi^2\chi^2$ of the inflaon field
$\phi$ with another scalar field $\chi$.

Furthermore, in these models since the inflaton field can not decay away efficiently,
its energy density must decrease extremely quickly
(${\rm e.g.},~ \rho_{\phi}\propto a^{-6}$)
soon after the end of inflation so that it cannot dominate the universe
again too early. Otherwise, the successful achievements of big-bang nucleosynthesis (BBN), CMB and
structure formations would be spoiled. This rapid diminishing
phase was called ``kination'' in Ref. \cite{joyce} or ``deflation'' in Ref.
\cite{no}.
It requires a steep region in the potential connecting to the flat
plateau for inflation. This often leads to the potential being strongly curved near
the end of inflation and tend to result in a power spectrum too far from
scale-invariance in some NO models. This problem is called $\eta$-problem in the
literature \cite{dimo,liddle2}, because in these models the predicted spectral
indices
\be\label{index}
n_s=1+2\eta-6\epsilon~,
\ee
are in conflict with current observational value $n_s=0.93\pm 0.03$ \cite{wmap1,wmap2}
mainly due to a large $|\eta|$. In Eq.
(\ref{index}) the slow-roll parameters are defined as
\be
\epsilon\equiv
\frac{M_{pl}^2}{16\pi}(\frac{V'}{V})^2,~~~~\eta\equiv\frac{M_{pl}^2}{8\pi}\frac{V''}{V}~,
\ee
where $M_{pl}=1.22\times 10^{19}$GeV is the Planck mass and
prime denotes the derivative with respect to $\phi$.
In addition, in a concrete NO model, the so-called original ``quintessentail inflation''
\cite{vilenkin} (which will be quoted as the example in present paper),
the inflation epoch is the same as
chaotic inflation with potential $V=\lambda \phi^4$. It predicts
a large gravitational wave amplitude, disfavored
by the WMAP data \cite{wmap3}.

In this paper we put forward another reheating mechanism for NO models by introducing
the curvaton mechanism \cite{lyth} which received many
attentions recently in the literature
\cite{dimo,liddle2,enqv,liddle,wands,moroi,enqvist}, and the problems
mentioned above can be ameliorated.
In our scenario, the reheating mechanism is based on the
oscillations and decays of another scalar field $\sigma$ (the
curvaton field) which is sub-dominant during inflation and has no interactions with inflaton except
the gravitational coupling. This process is as same as that of the
standard reheating in the usual inflation models.
We will show by an example that the reheating temperature
in our case can be as high as $10^{13}$GeV.
In addition, in the curvaton mechanism the curvature (adiabatic) perturbations are not originated
from the fluctuations of the inflton field, but instead from those of curvaton \cite{lyth}.
This happens in two steps. First, the quantum fluctuations in the
curvaton field during inflation are converted into classical
perturbations, corresponding to isocurvature perturbations when
cosmological scales leave the horizon. Then,
after inflation when the energy density of the curvaton
field becomes significant, the isocurvature perturbations in it
are converted into sizable curvature ones \cite{mollerach}. The
curvature perturbations become pure when the curvaton field or
the radiation it decays into begins to dominate the universe. Because the
curvaton is very light during inflation, the final spectrum of the curvature
perturbations is nearly flat and naturally consistent with current observations.
At the same time, in curvaton
scenario the fluctuations in inflaton field is negligible, so the
energy scale of inflation is relatively lower and the
gravitational wave amplitude is very small. Existing observations
have detected no gravitational waves, hence
those inflation models which usually predicted unlikely large tensor perturbations based on
the standard paradigm of inflaton-generating curvature perturbations,
like $\lambda \phi^4$ (and the example NO model cited below) \cite{wmap3},
would become viable in the curvaton scenario.
The primordial density
perturbations in curvaton scenario have been studied specifically in
Ref. \cite{wands} and its effects on CMB analyzed by the authors
of Ref. \cite{moroi,lewis}. Also, the curvaton mechanism has been used
to liberate some inflation models \cite{liddle,dimo}. In addition, there were some motivations for
the curvaton model from particle physics \cite{lyth,enqvist}.

For a specific presentation, we consider the quintessential inflation model
given by Peebles and Vilenkin \cite{vilenkin}.
In this model, the inflaton field will survive to
now to drive the present accelerated expansion suggested by recent measurements of type Ia
supernova \cite{super}. So, it can not decay to other particles. The potential of the infaton field is:
\bea
V(\phi)&=&\lambda(\phi^4+M^4)~~~{\rm for}~~\phi<0\nonumber\\
&=&\frac{\lambda M^8}{\phi^4+M^4}~~~~~~{\rm for}~~\phi\geq 0~.
\eea
The energy scale $M$ is assumed very small compared with the Planck mass,
the requirement of $\phi$ driving the present accelerated expansion
gives that $\phi_0>M_{pl}$ and $M>10^5$GeV, the
subscript $0$ represents present value.
The inflation which happened at $-\phi\gg M$ is chaotic type with $V=\lambda\phi^4$.
The slow-roll parameters are
\be
\epsilon=\frac{M_{pl}^2}{\pi
\phi^2}~,~~~\eta=\frac{3M_{pl}^2}{2\pi\phi^2}~.
\ee
Inflation ends when $\epsilon=1$, this corresponds to
$\phi_{end}\simeq -0.56 M_{pl}$. The e-folding number between the
horizon-exit of the comoving mode and the termination of inflation
can be estimated by
\be
N\equiv \int^{t_{end}}_{t_N}Hdt\simeq
\frac{\pi\phi^2_{N}}{M_{pl}^2}-1~.
\ee
Hence,
\be
\phi_N\simeq -\sqrt{\frac{N+1}{\pi}}M_{pl}~,
\ee
and
\be
\epsilon(\phi_N)\simeq \frac{1}{N+1}~.
\ee
The power spectra of generated curvature and tensor perturbations by inflaton are
respectively
\cite{book},
\bea
{\cal P}_{\zeta}=\frac{8V}{3M_{pl}^4 \epsilon }~,\nonumber\\
{\cal P}_{h}=\frac{128V}{3M_{pl}^4}~.
\eea
If the cosmic density perturbations of order $10^{-10}$ rely completely on this
curvature perturbation, the tensor to curvature ratio $r\equiv {\cal P}_{h}/{\cal
P}_{\zeta}=16\epsilon$ will be too large (for $N=50$, $r=0.32$) to
be compatible with WMAP data at $3-\sigma$ \cite{wmap3}.

This difficulty can be alleviated in the curvaton mechanism.
In this mechanism the curvature perturbations produced by the
inflaton is supposed to be negligible in contrast with COBE data ${\cal P}^{1/2}\sim 4.8\times 10^{-5}$.
This leads to
\bea
& &\lambda< 10^{-14}~,\nonumber\\
& &H_{in}<10^{-6}M_{pl}~,
\eea
where the subscript $in$ denotes the value taken by the corresponding variable in the
inflationary epoch. So, the gravitational wave amplitude,
proportional to $H_{in}/M_{pl}$, is negligibly small. In the
original paper of curvaton scenario, only a part of the components
of the universe come from the decays of the curvaton field. In
our case, since inflaton could not decay, we can assume that the
whole population of the universe is produced by curvaton. We will
show that this process can be very efficient.

We choose the quadratic potential for the curvaton field,
$V(\sigma)=m^2\sigma^2/2$, it has a minimum in the zero point of
$\sigma$ and $m\ll H_{in}$.

\begin{figure*}
\includegraphics[scale=.4]{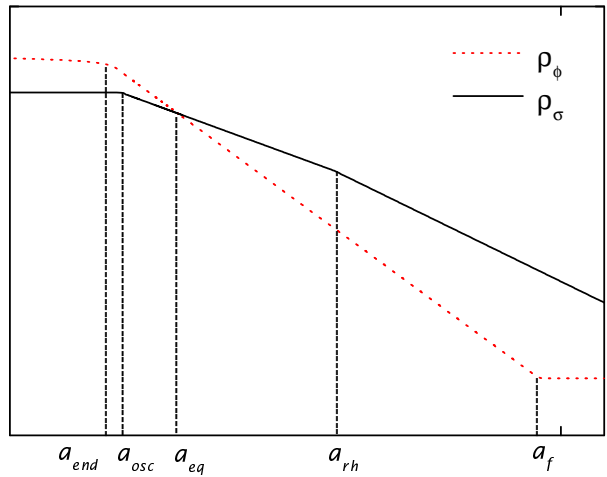}
\caption{\label{fig1} Evolution of the energy densities in the inflaton field (dotted line)
and the curvaton field (solid line). The subscripts `end', `osc',
`eq', `rh', and `f' respectively refer to the end of inflation,
the onset of curvaton oscillation, the equality of the energy
densities of $\phi$ and $\sigma$, reheating and freezing of
$\phi$.}
\end{figure*}

We numerically studied the evolution of the energy densities in the inflton
field and in the curvaton field. As demonstrated in Fig.
\ref{fig1}, in the era of inflation, the curvaton field slowly
rolled towards the minimum along its potential and its energy density changed little. Inflation ended at
$a=a_{end}$ when $\phi\simeq -0.5M_{pl}$, $a$ is the scale factor.
Then the universe entered into the kination period dominated by the kinetic
energy of inflaton,
during which:
\bea
\rho_{\phi}\propto a^{-6}~,~~H\propto a^{-3}~.
\eea
The curvaton field held slow-rolling at the beginning of
kination because of $m\ll H$, and its energy density
$\rho_{\sigma}$ decreased very slowly. Then it began to
oscillate about its minimum at $a=a_{osc}$ when $H$ decreased below $m$.
To avoid another inflation stage driven by the curvaton field,
the universe must still be dominated by $\rho_{\phi}$ at this point, this
requires
\be
\rho_{\sigma osc}<\rho_{\phi osc}=\frac{3H_{osc}^2 M_{pl}^2}{8\pi}~.
\ee
From
\be\label{osc}
H_{osc}\simeq m~, ~~ \rho_{\sigma osc}\simeq m^2\sigma_{osc}^2~,
\ee
we can estimate that:
\be\label{sigma}
\sigma_{osc}^2<\frac{3M_{pl}^2}{8\pi}.
\ee
After the curvaton field began oscillating, $\rho_{\phi}$
continued to decrease as $a^{-6}$ towards the value of present
energy density of dark energy and subsequently froze almost like a
cosmological constant when $a\geq a_f$.
$\rho_{\sigma}$ red-shifted as $a^{-3}$ like that of matter when averaged over many oscillations
until it decayed into radiation at $a=a_{rh}$, the subscript $rh$
means reheating. This happened when $H=H_{rh}\sim \Gamma$ and
$\Gamma$ is the decay rate of the curvaton field (we have assumed
instant reheating). It can occur after the moment when the
curvaton field dominated the universe ($a_{rh}\geq a_{eq}$) or at
the epoch when the curvaton field was still sub-dominant
($a_{rh}<a_{eq}$).

In the first case, as shown in Fig. \ref{fig1}, this requires:
\be
\rho_{\phi rh}\leq \rho_{\sigma rh}~.
\ee
Since $\rho_{\phi}\propto a^{-6}$, $\rho_{\sigma}\propto a^{-3}$
and
\be
H_{osc}/H_{eq}=(a_{eq}/a_{osc})^3~,
~~~~H_{eq}^2/H_{rh}^2=(a_{rh}/a_{eq})^3~,
\ee
we get
\be
\frac{\Gamma^2}{mH_{eq}}\leq \frac{8\pi
\sigma_{osc}^2}{3M_{pl}^2}~.
\ee
In above we have used Eq. (\ref{osc}). From $\rho_{\sigma
eq}=\rho_{\phi eq}$
we have
\be
H_{eq}=\frac{8\pi m \sigma_{osc}^2}{3M_{pl}^2}~,
\ee
so the constraint is (with Eq. (\ref{sigma})):
\be
\frac{\Gamma}{m}\leq \frac{8\pi\sigma_{osc}^2}{3M_{pl}^2}<1~.
\ee
The reheating temperature can be estimated from:
\be
\rho_{\sigma rh}=\frac{3M_{pl}^2}{8\pi}\Gamma^2\simeq
\rho_{rad}=\frac{\pi^2}{30}g_{\ast}T_{rh}^4~,
\ee
where $g_{\ast}$ counts the total number of
degrees of freedom of the relativistic particles, $g_{\ast}=10.75$ at temperature $T\sim
1$MeV, and for $T\geq 100$GeV, $g_{\ast}\sim {\cal O}(100)$.
The reheating process must be completed before BBN,
so in this case, the reheating temperature is approximately
\be\label{reheat1}
T_{rh}\simeq 0.78g_{\ast}^{-1/4}\sqrt{M_{pl}\Gamma}\leq
\sqrt{m\sigma_{osc}^2/M_{pl}}~.
\ee

Similarly, in the latter case, when the curvaton field decayed
when it was still sub-dominant but later than oscillation
($a_{osc}\leq a_{rh}<a_{eq}$), we have (for general, we assume
$\sigma_{osc}>0$)
\bea\label{reheat2}
& &\frac{8\pi\sigma_{osc}^2}{3M_{pl}^2}<\frac{\Gamma}{m}<1~,\nonumber\\
&
&\sqrt{m\sigma_{osc}^2/M_{pl}}<T_{rh}<0.4\sqrt{m\sigma_{osc}}~.
\eea
In this case, the radiation dominated era began with $T_{eq}\sim
(m\sigma^2)^{3/4}/(\Gamma^{1/4}M_{pl})$ which is in the range
\be
\sqrt{m\sigma_{osc}^3}/M_{pl}<T_{eq}<0.6 \sqrt{m\sigma_{osc}^2/M_{pl}}~.
\ee

For evaluation of $T_{rh}$, one has to consider the constraints on the mass of $\sigma$ field
and its initial value when it began oscillating, $\sigma_{osc}$.
Based on the curvaton mechanism \cite{lyth}, the resulted power
spectrum of the final curvature perturbations and its index are given by
\bea
& &{\cal P}^{1/2}_{\zeta}=\frac{H_{in}}{3\pi\sigma_{in}}\simeq
0.1\frac{H_{in}}{\sigma_{in}}~,\\
& &n_s\equiv 1+d\ln {\cal P}_{\zeta}/d\ln
k=1-2\epsilon+2m^2/(3H^2_{in})~.
\eea
One can see that the spectrum is nearly
flat. The curvaton field rolled very slowly before oscillation, we
can evaluate that $\sigma_{osc}\sim \sigma_{in}$.
Combined these results with the COBE normalization
and the limit $H_{in}< 10^{-6}M_{pl}$, one
has
\be\label{sigma2}
\sigma_{osc}< 10^{-3}M_{pl}\sim 10^{16}{\rm GeV}~.
\ee
Substituting Eq. (\ref{sigma2}) into Eq. (\ref{reheat1}) and
(\ref{reheat2}), and considering $m\ll H_{in}$ (we choose $m\leq
10^{-7}M_{pl}\sim 10^{12}$GeV for an evaluation), in both cases we
have approximately
\be
T_{rh}\lesssim 10^{13}{\rm GeV}~.
\ee

To ensure above analytical estimations  we have made a simple
numerical calculation. We set $\lambda=10^{-15}$, $M=4.7\times 10^{-14} M_{pl}$,
$m_{\sigma}=10^{-8}M_{pl}$ and $\sigma_{in}=3.27\times 10^{-3}
M_{pl}$.
For curvaton decays at $a=a_{eq}$, one gets $T_{rh}\sim 1.8\times 10^{12}$GeV. Assuming that
 curvaton decays into two fermions $\sigma\rightarrow f \overline{f}$, the decay width takes
the form $\Gamma= g^2 m_{\sigma}/8 \pi$. We find for $\sigma$ decays at $a=a_{eq}$ one requires
$g\sim 0.05 $, which is quite resonable. When the curvaton decays at $a<a_{eq}$, $g<1$ gives
$T_{rh}< 8\times 10^{12}$GeV. The BBN constraint that $T_{rh}>1$MeV
can also be easily satisfied, as from Fig. \ref{fig1}.

In conclusion, we have introduced the curvaton mechanism into the
non-oscillatory inflationary models as a possible reheating mechanism and did specific
studies in the quintessential inflation model proposed by Peebles and
Vilenkin. In our scenario, the reheating process is not attributed to the gravitational particle
productions, but due to the late-decaying of the curvaton field.
This mechanism, as same as the standard reheating mechanism which based on the oscillations and decays of
the scalar field (inflaton) in the usual chaotic inflation models,
can be efficient. Our calculations showed that the reheating temperature
can be as high as $10^{13}$GeV, it is high enough for the
quintessential baryogenesis \cite{li,trodden}.
At the same time, in addition to reheat the universe,
curvaton has been endowed the role of generating curvature perturbations
as in the original paper \cite{lyth}. By introducing curvaton mechanism into the NO models, the
predicted power spectra of curvature perturbations are nearly scale-invariance
and the tensor perturbations are negligibly small, consistent with current observations.
These properties can ameliorate other defects owned by some NO models. For
example, in Peebles-Vilenkin model we quoted in this paper,
the inflationary period is as same as the chaotic inflation
$V=\lambda \phi^4$, it predicts unlikely large gravitational
waves according to the usual paradigm of inflaton-generating
curvature perturbations \cite{wmap3}. For some other NO models,
the density perturbations are far from scale-invariance \cite{dimo,liddle2},
in conflict with observations if the
cosmic inhomogeneities are only due to the fluctuations of the
inflaton field. Our scenario provides a picture that the particle productions and the cosmic
density perturbations are both due to the curvaton field, this made it different from the original curvaton
scenario (in which only a part of the components of the universe come from curvaton) and other
reheating mechanisms that had been brought forward for NO inflationary models. Our work has also provided
a new way to the reheating problem of tachyon inflation \cite{tachyon}.

While this work was in progress a related paper \cite{dimo} appeared.
In Ref. \cite{dimo} the author mainly dealt with qunintessential inflation model building and parameter
restriction and our paper is focusing on the curvaton reheating mechanism.

{\bf{Acknowledgments:}} We would like to thank Prof. Xinmin Zhang and Dr. Yun-Song Piao for
useful discussions. We also thank Prof. David Lyth and Prof. David Wands for correspondences.
This work is supported in part by National Natural
Science Foundation of China and by Ministry of Science and Technology of China
under Grant No. NKBRSF G19990754.

{}

\end{document}